\documentstyle[twoside]{article}

\catcode`\@=11
\long\def\@makefntext#1{
\protect\noindent \hbox to 3.2pt {\hskip-.9pt  
$^{{\eightrm\@thefnmark}}$\hfil}#1\hfill}		

\def\thefootnote{\fnsymbol{footnote}}
\def\@makefnmark{\hbox to 0pt{$^{\@thefnmark}$\hss}}	
	
\def\ps@myheadings{\let\@mkboth\@gobbletwo
\def\@oddhead{\hbox{}
\rightmark\hfil\eightrm\thepage}   
\def\@oddfoot{}\def\@evenhead{\eightrm\thepage\hfil
\leftmark\hbox{}}\def\@evenfoot{}
\def\sectionmark##1{}\def\subsectionmark##1{}}

\oddsidemargin=\evensidemargin
\renewcommand{\thefootnote}{\fnsymbol{footnote}}

\newcounter{sectionc}\newcounter{subsectionc}\newcounter{subsubsectionc}
\renewcommand{\section}[1] {\vspace{12pt}\addtocounter{sectionc}{1} 
\setcounter{subsectionc}{0}\setcounter{subsubsectionc}{0}\noindent 
	{\tenbf\thesectionc. #1}\par\vspace{5pt}}
\renewcommand{\subsection}[1] {\vspace{12pt}\addtocounter{subsectionc}{1} 
	\setcounter{subsubsectionc}{0}\noindent 
	{\bf\thesectionc.\thesubsectionc. {\kern1pt \bfit #1}}\par\vspace{5pt}}
\renewcommand{\subsubsection}[1] {\vspace{12pt}\addtocounter{subsubsectionc}{1}
	\noindent{\tenrm\thesectionc.\thesubsectionc.\thesubsubsectionc.
	{\kern1pt \tenit #1}}\par\vspace{5pt}}
\newcommand{\nonumsection}[1] {\vspace{12pt}\noindent{\tenbf #1}
	\par\vspace{5pt}}

\newcounter{appendixc}
\newcounter{subappendixc}[appendixc]
\newcounter{subsubappendixc}[subappendixc]
\renewcommand{\thesubappendixc}{\Alph{appendixc}.\arabic{subappendixc}}
\renewcommand{\thesubsubappendixc}
	{\Alph{appendixc}.\arabic{subappendixc}.\arabic{subsubappendixc}}

\renewcommand{\appendix}[1] {\vspace{12pt}
        \refstepcounter{appendixc}
        \setcounter{figure}{0}
        \setcounter{table}{0}
        \setcounter{lemma}{0}
        \setcounter{theorem}{0}
        \setcounter{corollary}{0}
        \setcounter{definition}{0}
        \setcounter{equation}{0}
        \renewcommand{\thefigure}{\Alph{appendixc}.\arabic{figure}}
        \renewcommand{\thetable}{\Alph{appendixc}.\arabic{table}}
        \renewcommand{\theappendixc}{\Alph{appendixc}}
        \renewcommand{\thelemma}{\Alph{appendixc}.\arabic{lemma}}
        \renewcommand{\thetheorem}{\Alph{appendixc}.\arabic{theorem}}
        \renewcommand{\thedefinition}{\Alph{appendixc}.\arabic{definition}}
        \renewcommand{\thecorollary}{\Alph{appendixc}.\arabic{corollary}}
        \renewcommand{\theequation}{\Alph{appendixc}.\arabic{equation}}
        \noindent{\tenbf Appendix \theappendixc #1}\par\vspace{5pt}}
\newcommand{\subappendix}[1] {\vspace{12pt}
        \refstepcounter{subappendixc}
        \noindent{\bf Appendix \thesubappendixc. {\kern1pt \bfit #1}}
	\par\vspace{5pt}}
\newcommand{\subsubappendix}[1] {\vspace{12pt}
        \refstepcounter{subsubappendixc}
        \noindent{\rm Appendix \thesubsubappendixc. {\kern1pt \tenit #1}}
	\par\vspace{5pt}}

\newcommand{\textlineskip}{\baselineskip=13pt}
\newcommand{\smalllineskip}{\baselineskip=10pt}

\def\eightcirc{
\begin{picture}(0,0)
\put(4.4,1.8){\circle{6.5}}
\end{picture}}
\def\eightcopyright{\eightcirc\kern2.7pt\hbox{\eightrm c}} 

\newcommand{\copyrightheading}[1]
	{\vspace*{-2.5cm}\smalllineskip{\flushleft
	{\footnotesize International Journal of Modern Physics D, #1}\\
	{\footnotesize $\eightcopyright$\, World Scientific Publishing
	 Company}\\
	 }}

\newcommand{\publisher}[2]{{\begin{center}\footnotesize\smalllineskip 
	Received #1\\
	Revised #2
	\end{center}
	}}

\def\abstracts#1#2#3{{
	\centering{\begin{minipage}{4.5in}\baselineskip=10pt\footnotesize
	\parindent=0pt #1\par 
	\parindent=15pt #2\par
	\parindent=15pt #3
	\end{minipage}}\par}} 

\newcommand{\bibit}{\nineit}

\renewenvironment{thebibliography}[1]
        {\frenchspacing
	 \ninerm\baselineskip=11pt
         \begin{list}{\arabic{enumi}.}
        {\usecounter{enumi}\setlength{\parsep}{0pt}     
	 \setlength{\leftmargin 12.7pt}{\rightmargin 0pt} 
         \setlength{\itemsep}{0pt} \settowidth
	{\labelwidth}{#1.}\sloppy}}{\end{list}}

\newcounter{itemlistc}
\newcounter{romanlistc}
\newcounter{alphlistc}
\newcounter{arabiclistc}

\newcommand{\fcaption}[1]{
        \refstepcounter{figure}
        \setbox\@tempboxa = \hbox{\footnotesize Fig.~\thefigure. #1}
        \ifdim \wd\@tempboxa > 5in
           {\begin{center}
        \parbox{5in}{\footnotesize\smalllineskip Fig.~\thefigure. #1}
            \end{center}}
        \else
             {\begin{center}
             {\footnotesize Fig.~\thefigure. #1}
              \end{center}}
        \fi}

\newcommand{\tcaption}[1]{
        \refstepcounter{table}
        \setbox\@tempboxa = \hbox{\footnotesize Table~\thetable. #1}
        \ifdim \wd\@tempboxa > 5in
           {\begin{center}
        \parbox{5in}{\footnotesize\smalllineskip Table~\thetable. #1}
            \end{center}}
        \else
             {\begin{center}
             {\footnotesize Table~\thetable. #1}
              \end{center}}
        \fi}

\def\@citex[#1]#2{\if@filesw\immediate\write\@auxout
	{\string\citation{#2}}\fi
\def\@citea{}\@cite{\@for\@citeb:=#2\do
	{\@citea\def\@citea{,}\@ifundefined
	{b@\@citeb}{{\bf ?}\@warning
	{Citation `\@citeb' on page \thepage \space undefined}}
	{\csname b@\@citeb\endcsname}}}{#1}}

\newif\if@cghi
\def\cite{\@cghitrue\@ifnextchar [{\@tempswatrue
	\@citex}{\@tempswafalse\@citex[]}}
\def\citelow{\@cghifalse\@ifnextchar [{\@tempswatrue
	\@citex}{\@tempswafalse\@citex[]}}
\def\@cite#1#2{{$\null^{#1}$\if@tempswa\typeout
	{IJCGA warning: optional citation argument 
	ignored: `#2'} \fi}}

\def\pmb#1{\setbox0=\hbox{#1}
	\kern-.025em\copy0\kern-\wd0
	\kern.05em\copy0\kern-\wd0
	\kern-.025em\raise.0433em\box0}

\def\fnt#1#2{\footnotetext{\kern-.3em
	{$^{\mbox{\scriptsize #1}}$}{#2}}}

\def\fpage#1{\begingroup
\voffset=.3in
\thispagestyle{empty}\begin{table}[b]\centerline{\footnotesize #1}
	\end{table}\endgroup}

\font\tenrm=cmr10
\font\tenit=cmti10 
\font\tenbf=cmbx10
\font\bfit=cmbxti10 at 10pt
\font\ninerm=cmr9
\font\nineit=cmti9

\font\eightrm=cmr8

\def\qed{\hbox{${\vcenter{\vbox{			
   \hrule height 0.4pt\hbox{\vrule width 0.4pt height 6pt
   \kern5pt\vrule width 0.4pt}\hrule height 0.4pt}}}$}}

\renewcommand{\thefootnote}{\fnsymbol{footnote}}	

\begin{document}

\normalsize\textlineskip
\thispagestyle{empty}
\setcounter{page}{1}

\copyrightheading{}			

\vspace*{0.88truein}

\fpage{1}
\centerline{\bf QUANTUM SQUEEZING AND LATE TIME CLASSICAL BEHAVIOR OF }
\vspace*{0.035truein}
\centerline{\bf MASSIVE FIELDS IN EXPANDING ROBERTSON-WALKER UNIVERSE}
\vspace*{0.37truein}
\centerline{\footnotesize Milan Miji\'c\footnote{e-mail address: milan@moumee.
calstatela.edu}}
\vspace*{0.015truein}
\centerline{\footnotesize\it Department of Physics and Astronomy, 
California State University Los Angeles}
\baselineskip=10pt
\centerline{\footnotesize\it Los Angeles, CA 90032,}
\vspace*{0.015truein}
\centerline{\footnotesize\it and Institute for Physics, P.O. Box 522,}
\baselineskip=10pt
\centerline{\footnotesize\it 11001 Belgrade, Yugoslavia}
\vspace*{0.225truein}
\publisher{}{}

\vspace*{0.21truein}
\abstracts{A Schroedinger picture analysis of time dependent quantum 
oscillators, in a manner of Guth and Pi, clearly identifies 
{\it two} physical mechanisms for possible decoherence of vacuum fluctuations 
in early universe: turning of quantum oscillators upside-down, and rapid 
squeezing of upside-right oscillators so that certain squeezing factor 
diverges. In inflationary cosmology the former mechanism explains the 
stochastic evolution of light inflatons and the classical nature of density 
perturbations in most of inflationary models, while the later one is 
responsible for the classical evolution of relatively heavy fields, with 
masses in a narrow range above the Hubble parameter: 
$m/H_0 \in (\sqrt {2}, 3/2)$. 
The same method may be applied to study of the decoherence of quantum
fluctuations in any Robertson-Walker cosmology.}{}{}


\textlineskip			
\vspace*{12pt}			

\vspace*{-0.5pt}
\noindent
\textheight=7.8truein
\setcounter{footnote}{0}
\renewcommand{\thefootnote}{\alph{footnote}}

\section{Introduction} 

\noindent
One of the great achievements of inflationary cosmology,$^{1,2}$ 
is the observation$^3$ that quantum fluctuations 
of inflaton field evolve into classical adiabatic density perturbations 
which might have served as seeds for the observed large scale structure of the
universe. The spectrum of these perturbations is very close to the
Harrison-Zeldovich spectrum, which agrees reasonably well with observations. 
With the current uncertainties in physics above the electro-weak scale, 
the magnitude of the perturbations is sufficiently
adjustable to avoid the conflict between the inflationary models and, say,
the observed anisotropy of the cosmic microwave background.$^2$
On a conceptual level however, this mechanism is an excellent example for 
spontaneous emergence of classical features in course of the evolution of a 
quantum universe. This is the angle that we are interested in here.

There have been several attempts over the last decade
to describe and understand this decoherence process.$^{4,5,6}$
Following the line of thinking due to Guth and Pi,$^4$
we formulate here sufficient conditions for the 
decoherence of quantum fluctuations of scalar fields in any expanding 
universe which is described by the Robertson-Walker geometry.

We start by classifying different possibilities, and stating results
known so far. Then, we analyze the time dependent oscillators, as model that 
adequately describes the quantum fluctuations of free fields in expanding 
universe, and formulate the conditions for decoherence.
Next, we apply this analysis to massive fields on De Sitter spacetime,
and distinguish and discuss the characteristic cases. The last section 
contains our conclusions  and some inevitable speculations.
As usual, we will neglect 
the effects of the interaction, due to the expected smallness of both the 
coupling constant(s) and the fluctuations themselvs. Moreover, one would
expect that if the decoherence shows at the zeroth order approximation,
an account for interaction ought to make it only better.

\bigskip
\section{Upside-down and Upside-right oscillators}

\noindent
Let $\delta \phi(\vec {x}, t)$ 
stands for inhomogeneous fluctuation of some scalar field we are interested
in. After expanding the field $\delta \phi$ into modes counted by the
wave number $k$, introducing the conformal time $\eta$ as $dt = S(\eta) 
d\eta, ~~S(\eta) = a(t)$, and the rescaled modes $\chi$ as $\delta \phi =
\chi/S(\eta) $, (the index $k$ will be suppressed), the one-mode equation
of motion takes the following simple form:

\begin{equation}
\chi^{\prime \prime}(\eta) + \omega^2 (\eta) \chi (\eta) = 0 ~~,
\label {osc}
\end{equation}
\noindent
with,
\begin{equation}
\omega^2 (\eta) = k^2 + m^2 S^2 - S^{\prime \prime}/S ~~.
\label {omega}
\end{equation}

Different expansion laws lead to different behavior of the one mode
frequency (\ref {omega}).
The most general characteristic of inflationary expansion is that it is
accelerated, $\ddot {a}(t) >0 $. This means that the dominant energy condition
is violated, and the physical wavelength of any mode will eventually exceed
the Hubble radius. The potential will turn upside-down, if, at
some moment, the curvature term exceeds the mass term in 
Eq. (\ref {omega}): $S^{\prime \prime}/S > m^2 S^2$. Introducing kinetic
function $v(S) \equiv S^{\prime 2}/2$, these two conditions may be
expressed as follows,

\medskip
\noindent
Inflation
\begin{equation}
\frac {v^{\prime}}{S} > 2 \frac {v}{S}~~.
\end{equation}
\noindent
Upside-down oscillator:
\begin{equation}
\frac {v^{\prime}}{S} > m^2 S^2 ~~.
\end{equation}

It is now easy to see that if the oscillator is upside-down, the background
must be inflationary, i.e., that the second condition implies the first.
From the analysis of Guth and Pi$^4$ we know that the late time
behavior of such oscillator is essentially classical. However, the opposite 
is not true. 

In De Sitter universe the shape of the oscillator is determined by the mass
of the field relative to the (constant) Hubble parameter. For 
$0 \leq m^2/H_0^2 < 2$, the oscillator turns upside-down when the physical 
wavelength of the mode crosses the Hubble radius or some time after that 
moment.
For $m^2 < 0$ the turning of the oscillator happens even before the Hubble 
radius crossing. Thus, those modes have essentially classical behavior 
at late times, and, according to Guth and Pi$^4$, as they re-enter the Hubble 
radius in subsequent non-inflationary phase they may be interpreted as 
(the source of) classical density perturbations. The situation is different
in case of heavy fields with $m^2/H_0^2 > 2$.  The fluctuations of those
fields also cross the Hubble radius, but their oscillators are upside-right
at all times. In this case the analysis of Guth and Pi$^4$ does not tell us 
anything about the possible classical behavior, or the lack of it.

The behavior of these, fairly massive fields, have been investigated in
Ref. 6, and led to somewhat unexpected result that the fluctuations of fields 
with mass in a narrow range $(\sqrt {2}, 3/2)$ times $H_0$ also decohere, 
although 
they are represented by upside-right oscillators. The fluctuations of heavier
fields were found not to decohere, even when their wavelengths are
greater than the Hubble radius.

In this paper we will show how one can confirm and understand all these 
possibilities in an elementary way.  By following the analysis of Guth and 
Pi,$^4$ we will find that for quantum oscillator in any expanding 
Robertson-Walker geometry there is not just one mechanism for 
decoherence found by them, (turning of the oscillator upside-down), but that 
there is also the second one. In De Sitter case those two mechanisms exactly 
cover the two possible regimes of decoherence quoted above.

\bigskip
\section{Time dependent oscillators}

\noindent
Let us consider a one-dimensional, time dependent oscillator
with the Hamiltonian $H = p^2/(2m) + k(t)x^2/2$. The frequency of such 
oscillator is given as $\omega^2(t) = k/m$. Let $x_c(t)$ denotes a 
solution to the classical equation of motion, 
$\ddot {x}_c + \omega^2(t) x_c =0$, and $p_c(t) \equiv m \dot {x}_c (t)$ its
conjugate momentum. Then, the solution to the time-dependent
Schroedinger equation may be written as,

\begin{equation}
\Psi (x, t) = A_0 \left | \frac {x_c(0)}{x_c(t)} \right |^{1/2} \times
\exp \left [ \frac {i}{2}\frac {p_c(t)}{x_c(t)} x^2 \right ] ~~.
\label {oscwf}
\end{equation}

In their analysis of upside-down oscillator Guth and Pi$^4$ use three
different but related criteria as signals for the emergence of classical 
behavior. We will examine here the two of them: (i) the emergence of classical 
momentum, and, (ii) the condition for relatively small magnitude of the 
canonical commutator.

To see how the first criteria works, recall that
for an arbitrary amplitude, acting on it by the momentum operator creates
a state which bears no reference to classical momentum. However, in case
of amplitude (\ref {oscwf}), we have,

\begin{equation}
p\Psi (x, t) = \frac {p_c(t)}{x_c(t)} x \Psi (x, t) ~~.
\label {pPsi}
\end{equation}

To examine the ratio between the two classical quantities on the right hand 
side, we can rewrite the classical solution as,$^7$

\begin{equation}
x(t) = A(t) x_0 + B(t) p_0~~.
\label {xc}
\end{equation}
\noindent
$x_0$ and $p_0$ are the initial values, and time dependent coefficients $A$ and
$B$ are given as linear combinations of two independent solutions
to the oscillator equation:

\begin{equation}
A(t) \equiv \frac {x_{1}(t) x_2^{\prime}(0) -
x_{1}^{\prime}(0) x_{2}(t)}
{W(0)} ~~,
\label {A}
\end{equation}
\begin{equation}
B(t) \equiv \frac {x_{1}(0) x_{2}(t) -
x_{1}(t) x_{2}(0)}
{W(0)} ~~,
\label {B}
\end{equation}
\begin{equation}
W(0) \equiv
{x_{1}(0) x_{2}^{\prime}(0) -
x_{1}^{\prime}(0) x_{2}(0)} ~~.
\end{equation}

One can eliminate one of the two constants to express the classical momentum
either as,
\begin{equation}
p_c(t) = m \frac {\dot A}{A} x_c(t) + m B \left [ \frac {\dot {B}}{B} - 
\frac {\dot A}{A}
\right ] p_0~~,
\label {pp}
\end{equation}
\noindent
or, as,
\begin{equation}
p_c(t) = m \frac {\dot B}{B} x_c(t) + m A \left [ \frac {\dot {A}}{A} -
\frac {\dot B}{B}
\right ] x_0~~.
\label {px}
\end{equation}

In cases when the last term on the right hand side is absent, or small compared
to the first, we have $p_c \sim x_c$. Then, we can rewrite Eq. (\ref {pPsi})
as,
\begin{equation}
p\Psi (x, t) = p_c(x) \Psi(x, t) ~~,
\end{equation}
\noindent
where $p_c(x) \equiv m (\dot {A}/A) x$, or, $p_c(x) \equiv m (\dot {B}/B) x$,
depending which of the two possible realizations for $p_c \sim x_c$ takes
place. $p_c(x)$ denotes the classical momentum that oscillator has at
point $x$ as it reaches it along the classical trajectory specified by the
initial conditions $(x_0, p_0)$. 
In contrast with the general quantum mechanical expression, the
average value for the square of the momentum is now the average of classical 
values:

\begin{equation}
\langle \Psi | p^2(t) |\Psi \rangle = \int dx~ |\Psi (x,t)|^2 p_c^2(x) ~~.
\label {average}
\end{equation}

This is exactly how would one calculate the average in the ensemble
where position and momentum are correlated according to classical equation. 
Let us now see when can this occur. The relation
$p_c \sim x_c$ holds exactly if, 
(i) $W_{AB} \equiv A \dot {B} - B \dot {A} =0$, 
(ii) $p_0 = 0$, or, (iii) $x_0 = 0$. We will exclude the first case, as it means
that $x_0$ and $p_0$ cannot be expressed through $x_c(t)$ and $p_c(t)$, ie,
that the classical solution cannot be run backward. The third case should also
be excluded, as, from (\ref {oscwf}), it implies $\Psi (x, 0) = 0$, which 
violates the requirement for finite normalization. We are left only with the 
second case, which implies a uniform initial distribution: $\Psi (x, 0) = A_0$. 
For a finite range of the coordinate $x$ this is acceptable, otherwise this 
case should also be excluded because the amplitude will not be normalisable.

It is more interesting if we consider the regime when $p_c \sim x_c$ holds only
approximately. For this we need either, 

\medskip
\noindent
(a) $x_c(t)$ large, with $\dot {A}/A$,
and $\dot {B}/B$ finite, and $W_{AB}/A$, $W_{AB}/B$, $x_0$, and $p_0$ all 
bounded; or, 

\noindent
(b) $\dot {A}/A$ and $\dot {B}/B$ are large, $x_c(t)$ is finite,
and, as in the preceding case, the last terms in (\ref {pp}) and 
(\ref {px}) are bounded.

\medskip
Guth and Pi$^4$ case corresponds to the first of those two
possibilities. For an upside-down oscillator with constant frequency
$\omega_0$ the two independent solutions may be chosen as $x_{1,2} = \exp [
\pm \omega_0 t ]$. Eq. (\ref {pp}) becomes,

\begin{equation}
p_c(t) = \left [ m \omega_0 \tanh (\omega_0 t) \right ] x_c(t) +
\frac {p_0}{\cosh (\omega_0 t)} ~~.
\end{equation}

At late times, $\omega_0 t >> 1$, we have $p_c \rightarrow m \omega_0 x_c$,
and, following the previous logic, the late time behavior of an 
upside-down oscillator is essentially classical.

Another way to understand this case is to simply notice that at late times
the classical solution is accurately described through just one exponential
mode, the growing one. Since the same is true for $p_c$ we have $p_c \sim
\omega_0 x_c$, and classical behavior emerges. In the opposite case of an 
upside-right oscillator with constant frequency, the fundamental solutions 
are periodic, and both must be retained at late times. As a result, the 
second terms in Eq.'s (\ref {pp}), (\ref {px}) can't be neglected.
This distinction between the cases when just one or both modes describe
the late time behavior carries on to time dependent oscillators.

Let us now consider the second criteria of Guth and Pi.$^4$ This one is 
useful because it gives us a characteristic
scale in the configuration space outside of which the evolution is
effectively classical. By acting with
operators $xp$ and $px$ on the wave function (\ref {oscwf}) we find that
both results approximately agree if,

\begin{equation}
\left | x^2 \frac {p_c(t)}{x_c(t)} \right | \gg 1 ~~.
\label {2crit}
\end{equation}

For example, if we again consider the upside-down oscillator with constant 
frequency, we find that it behaves nearly classically at large amplitudes,$^4$
$ x \gg (m \omega_0)^{-1/2}$. In case of a time-dependent oscillator
the decoherence scale will depend on time. In a $p_c \sim x_c$ regime, this 
condition becomes,

\begin{equation}
|x p_c(x)| \gg 1 ~~,
\end{equation}

\noindent
which is the resurrection of the uncertainty principle: the behavior
near given value of $x$ is approximately classical if $x$ is greater than 
the De Broglie wavelength for the classical momentum that corresponds to
that value of $x$.

Let us now summarize the steps to see whether particular time-dependent
oscillator has nearly classical behavior. First, we solve the oscillator
equation to find the two independent solutions. Next, we calculate the
coefficients $A$ and $B$. If one of the conditions (a) or
(b) above is full-filed, the behavior is essentially classical. Finally,
from Equation (\ref {2crit}) we find for which values of the amplitude
(as opposed to at what times) the classical behavior takes place. Of course,
along the way, possible late time dominance of just one solution will give us
advance clue about the emergence of classical regime. In the
next section we will apply this procedure to the simplest inflationary model.

\bigskip  
\section{De Sitter expansion}

\noindent
Let us start by listing a few well known results about the massive fields on 
exact, spatially flat De Sitter background. The convenient time-like variable
is $z \equiv - k \eta \in (0, \infty)$. The one mode frequency is,

\begin{equation}
\omega^2(z) = 1 - \frac {1}{z^2} \cdot \left ( \frac {m^2}{H_0^2} - 2
\right ) ~~.
\end{equation}

\noindent
The independent solutions $\chi_{1,2}$ are given in terms of Hankel 
functions:

\begin{equation}
\chi_{1,2}(z) = \frac {\sqrt {\pi}}{2} \sqrt {z} H_{\nu}^{(1,2)}(z) ~,~~ 
\nu^2 \equiv \frac {9}{4} - \frac {m^2}{H_0^2}~~.
\end{equation}

As before, the classical solution and its momentum will be expressed as

\begin{equation}
\chi_c = \alpha \chi_1(z) + \beta \chi_2 (z) ~~,
\end{equation}
\begin{equation}
p_c = - k \left [\alpha \chi_1^{\prime}(z) + \beta \chi_2^{\prime}(z) \right ]
~~,
\end{equation}

\noindent
where $\alpha$ and $\beta$ are two complex constants.

Since we are interested in the late time behavior of the wave function we
only need the late time behavior of the solution. The late time ($z \rightarrow
0^+$) behavior of the modes is well known:

\begin{equation}
\chi_1(z) = C_{\nu} z^{1/2 + \nu} + D_{\nu} z^{1/2 - \nu}~,
\end{equation}
\begin{equation}
\chi_2(z) = A_{\nu} z^{1/2 + \nu} + B_{\nu} z^{1/2 - \nu}~~,
\end{equation}

\noindent
with,

\begin{equation}
A_{\nu} = \frac {\pi^{1/2} e^{i\nu \pi}}{2^{\nu + 1} \Gamma (\nu + 1)}
~(1 + i \cot \nu \pi )~~,
\end{equation}
\begin{equation}
B_{\nu} = D_{\nu} = \frac {i e^{i\nu \pi}}{\pi^{1/2}} 2^{\nu - 1} \Gamma (\nu)
~~,
\end{equation}
\begin{equation}
C_{\nu} = \frac {\pi^{1/2} e^{i\nu \pi}}{2^{\nu + 1} \Gamma (\nu + 1)}
~(1 - i \cot \nu \pi ) ~~.
\end{equation}

Consider first the case when $\nu$ is real. Then, at small values of $z$ 
both independent solutions behave as $z^{1/2 - \nu}$, and, effectively, we
have just one mode. Note that for $\nu > 1/2$ the dominant mode diverges,
while for $0<\nu < 1/2$ it goes to zero, but slower than the other mode which
is proportional to $z^{1/2 + \nu}$. In either case, as $z \rightarrow 0^+$,

\begin{equation}
\chi_c(z) = (\alpha + \beta) B_{\nu} z^{1/2 - \nu} + {\cal O}(z^{2\nu})~,
~~\nu \in {\cal R} ~,
\end{equation}
\noindent
and, it follows that,

\begin{equation}
p_c = (-)k \left ( \frac {1}{2} - \nu \right ) ~\frac {\chi_c}{z}~~.
\label {pvschi}
\end{equation}

This is the sufficient condition for the emergence of classical correlations.
The argument breaks down for $\nu = 1/2$, or, equivalently, $m^2 = 2 H_0^2$.
Minimally coupled field of that mass is equivalent to conformally coupled
massless field.

The situation is different when $\nu$ is imaginary. In that case we have,

\begin{equation}
\chi_c = z^{1/2}~ \left [ (\alpha C_{\nu} + \beta A_{\nu}) e^{i |\nu| \log z}
+ (\alpha + \beta) B_{\nu} e^{- i |\nu| \log z} \right ] ~~,
\end{equation}
\noindent
and 
\begin{equation}
p_c = z^{-1/2}~ \left [ \left (\frac {1}{2} + \nu \right )
(\alpha C_{\nu} + \beta A_{\nu}) e^{i |\nu| \log z}
+ \left ( \frac {1}{2} - \nu \right )
(\alpha + \beta) B_{\nu} e^{- i |\nu| \log z} \right ] ~~.
\end{equation}
\noindent
Both modes are oscillatory and grow or decay at the same rate. As both of them
must be retained, $p_c$ is not simply proportional to
$\chi_c$, and the classical correlation does not emerge.

Let us now try to see this through the behavior of the squeezing coefficients.
The initial conditions are assigned at $z=z_i \gg 1$, where, for
all values of $\nu$, modes behave as simple exponentials:

\begin{equation}
\chi_{1,2}(z_i) = \frac {1}{\sqrt{2}}\cdot e^{ \mp i \Phi }~,~~
\Phi \equiv z_i - \nu \frac {\pi}{2} - \frac {\pi}{4} ~.
\end{equation}
 
The squeezing coefficients $A$ and $B$  may then be calculated as,

\begin{equation}
A(z) = \frac {1}{\sqrt {2}} \cdot \left [ \chi_1(z) e^{i\Phi} +
\chi_2 (z) e^{-i \Phi} \right ] ~~,
\end{equation}
\begin{equation}
B(z) = \frac {i}{\sqrt {2}} \cdot \left [ \chi_1(z) e^{i\Phi} -
\chi_2 (z) e^{-i \Phi} \right ] ~~.
\end{equation}

We find that $\dot A = - k A^{\prime}(z)$, and similarly for the coefficient
$B$. All these expressions differ just in the constant coefficients from
those for the classical solution. Again, 
when $\nu$ is real, $z^{1/2 - \nu}$ mode dominates both $A$ and $B$,
and we have $A^{\prime}(z)
\sim A(z)$ and $B^{\prime} (z) \sim B(z)$. One therefore finds at $z \ll 1$,

\begin{equation}
\frac {\dot {A}}{A} = \frac {\dot {B}}{B}
= (-) k~ \left ( \nu - \frac {1}{2} \right ) \frac {1}{z}
~~,
\end{equation}

\noindent
and,
\begin{equation}
\dot {A} B - A \dot {B} = 0~~,
\end{equation}

\noindent
with ${\cal O}(z^{2\nu})$ corrections in both equations. Therefore, the
second term
in both (\ref {pp}) and (\ref {px}) behaves as $A \cdot W_{AB} \sim
B\cdot W_{AB} \sim {\cal O}(z^{1/2 + \nu})$. To estimate the first term
recall that the classical 
solution diverges as $\chi_c \sim z^{1/2 -\nu}$ for $\nu > 1/2$, and
decays to zero with the same rate if $\nu < 1/2$. The first terms in
(\ref {pp}, \ref {px}) behaves as $z^{-1} \cdot z^{1/2 - \nu}$ which diverges
for both ranges of $\nu$. The classical regime emerges in both cases,
but while for $\nu > 1/2$ it is due to the unbounded growth of the classical
solution in the upside-down potential, in case of $\nu < 1/2$ the classical
solution is bounded and the decoherence takes place due to the divergence
of squeezing coefficients ${\dot A}/A$ or ${\dot B}/B$. These are exactly
the two possibilities labeled as (a) and (b) in the preceding section.

In contrast, when $\nu$ is imaginary, we find,

\begin{equation}
\frac {\dot {A}}{A}  = (-) \frac {k}{z} 
\frac 
{ M_{\nu} (1/2 + \nu) e^{i |\nu| \log z} +  
N_{\nu} (1/2 - \nu) e^{- i |\nu| \log z}}
{ M_{\nu} e^{i |\nu| \log z} +  
N_{\nu} e^{- i |\nu| \log z}}~~,
\end{equation}
\noindent
with,
\begin{equation}
M_{\nu} \equiv C_{\nu} e^{i\Phi} + A_{\nu} e^{-i\Phi}~,
\end{equation}
\begin{equation}
N_{\nu} \equiv D_{\nu} e^{i\Phi} + B_{\nu} e^{-i\Phi}~.
\end{equation}

The expression for the rate of change of $B$ differs only in coefficients:

\begin{equation}
\frac {\dot {B}}{B}  = (-) \frac {k}{z} 
\frac 
{ \tilde {M}_{\nu} (1/2 + \nu) e^{i |\nu| \log z} +  
\tilde{N}_{\nu} (1/2 - \nu) e^{- i |\nu| \log z}}
{ \tilde {M}_{\nu} e^{i |\nu| \log z} +  
\tilde {N}_{\nu} e^{- i |\nu| \log z}}~~,
\end{equation}
\noindent
with,
\begin{equation}
\tilde {M}_{\nu} \equiv (-) C_{\nu} e^{i\Phi} + A_{\nu} e^{-i\Phi}~,
\end{equation}
\begin{equation}
\tilde {N}_{\nu} \equiv (-) D_{\nu} e^{i\Phi} + B_{\nu} e^{-i\Phi}~.
\end{equation}

The fractions in both expressions are bounded, so both $\dot {A}/A$ and
$\dot {B}/B$ diverge as $z^{-1}$. However, we must also check the magnitude of
the second term in Eq.'s (\ref {pp} - \ref {px}). Both modes behave as
$z^{1/2}$, and the same as true for $\chi_c$, $A$, and $B$. Thus, the first
terms in (\ref {pp} - \ref {px}) behave as $z^{-1/2}$. The second terms have
the same growth at late times, $ \sim z^{1/2} z^{-1} \sim z^{-1/2}$. Thus,
in general, the first term does not dominate. 

The last hope is that the second term vanishes
altogether, due to the exact cancellation, $\dot {A}/A = \dot {B}/B$. 
For this we need 
$(M_{\nu}, N_{\nu}) = (\tilde {M}_{\nu}, \tilde {N}_{\nu})$, which itself is
possible only if $C_{\nu} = D_{\nu} = 0$. (And hence $B_{\nu} = 0$. This
is reminiscent of the condition for the dominance of one mode.) For this
to happen we need $|\nu| \rightarrow \infty$, (hence $m^2/H_0 \rightarrow
\infty$ as well), but then also $A_{\nu} = 0$, so the only solution that we 
can represent in this way is $\chi_c = 0$. Therefore, we find that
neither of the conditions for decoherence is satisfied, and there is no
emergence of classical correlations for $\nu^2 < 0$.

Having now established the late time classical behavior of fluctuations
with mass smaller than $3H_0/2$ ($m^2 = 2 H_0^2$ is excluded, as it
corresponds to conformally coupled massless field), we can now use the
commutator criteria, Eq. (\ref {2crit}), to find out the range of amplitudes
which may be considered classical. Using Eq. (\ref {pvschi}), we find for
all real $\nu \not = 1/2$, that the behavior is classical for amplitudes,

\begin{equation}
|\chi^2| \gg \frac {|\eta|}{|\nu -1/2|}~.
\end{equation}

One should keep in mind that the wave number $k$ is fixed, and that the
adopted ratio (\ref {pvschi}) applies only at the late times, $z \rightarrow
0^+$, or equivalently, $k|\eta| \ll 1$. In terms of the original variable 
characterizing the fluctuation, the bound is given as,

\begin{equation}
|\delta \phi | \gg \frac {H_0 |\eta|^{3/2}}{|\nu - 1/2|^{1/2}}~.
\end{equation}

The overall normalisation to Hubble parameter is common to all various
measures of decoherence,$^5$ while particular scaling with time, or
the wave number depends on details of the definition of that measure. For
instance, for modes outside the Hubble radius, the decoherence bound on
$\delta \phi$ is much smaller than the zero coupling 
limit of the coherence length derived by Brandenberger et al.$^5$ for
a model with two coupled massless fields.

\bigskip
\section{Conclusions}

\noindent
Stochastic dynamics of inflationary cosmology, and the generation of
classical density perturbations through vacuum fluctuations of inflaton field,
are normally studied and understood only in cases when the inflaton, or
its fluctuations, behave as very light fields, $m^2 \ll H^2$. This is usually 
motivated by the observational fact that the universe is only slightly
inhomogeneous, which, in simplest models, translates into small mass or
coupling of the inflaton field. However, inflationary models
are getting more complex, scenarios are more complicated, and the relationship
between the observational quantities and parameters of the Lagrangean may
not be as straightforward. The emergence of classical spacetime in quantum
cosmology might be due to processes which have little or nothing to do with
the relics observed today. It is therefore important to know all 
possibilities for which important mechanisms, such as the generation of 
classical density perturbations, may take place. This paper develops 
a simple, but fairly general analysis of the time dependent quantum
oscillators, on basis of which one can formulate criteria for decoherence
of scalar field quantum fluctuations in any Robertson-Walker cosmology. The
method is just a simple extension of the approach taken by Guth and Pi,$^4$
but thanks to this extension one can confirm and understand results on 
decoherence in cases in which fluctuations are not described through upside-down
quantum oscillators. In particular, we confirm the results found in Ref. 6.

Finally, one should stress that the criteria of decoherence derived here
strictly apply to free fields. Normally one expects that the presence of a small
interaction ought to make decoherence stronger. More importantly, these
criteria are just sufficient, but not necessary conditions for decoherence.
Some other mechanism may bring decoherence in cases in which it does not
take place according to criteria used here. It is important to stress however,
that the mechanism described here appears to be sufficient to describe
the decoherence im most popular De Sitter-like inflationary models, as well
as in most of the models of power law inflation.$^8$ 

\nonumsection{References}

\end{document}